\documentclass[twocolumn]{revtex4-1}
\usepackage{amsmath}
\usepackage{amssymb}
\usepackage{bm}
\usepackage{epsfig}
\usepackage{graphicx}
\usepackage{color}
\usepackage[colorlinks = true]{hyperref}
\usepackage[T2A]{fontenc}
\usepackage[english]{babel}

\renewcommand{\phi}{\varphi}
\renewcommand{\kappa}{\varkappa}

\begin{document}

\newcount\timehh  \newcount\timemm
\timehh=\time \divide\timehh by 60
\timemm=\time
\count255=\timehh\multiply\count255 by -60 \advance\timemm by \count255

\title{Spin-dependent coherent transport of two-dimensional excitons}

\author{M.V. Durnev and M.M. Glazov}

\affiliation{Ioffe Institute, 194021 St.-Petersburg, Russia}

\date{\today, file = \jobname.tex, printing time = \number\timehh\,:\,\ifnum\timemm<10 0\fi \number\timemm}

\begin{abstract}
We propose a theory of interference contributions to the two-dimensional exciton diffusion coefficient. The theory takes into account four spin states of the heavy-hole exciton. An interplay of the single particle, electron and hole, spin splittings with the electron-hole exchange interaction gives rise to either localization or antilocalization behavior of excitons depending on the system parameters. Possible experimental manifestations of exciton interference are discussed.
\end{abstract}

\maketitle

\section{Introduction}

Excitonic effects play crucial role in optical properties of semiconductors and semiconductor nanosystems~\cite{excitons:RS,excitons:GS}. Exciton energy spectrum and wavefunctions determine the fine structure of optical absorption spectra and selection rules for optical transitions. Manipulations of excitonic states by external electric and magnetic fields, elastic strain, etc., pave way to control optical properties of semiconductors. That is why excitonic effects in bulk materials and in low-dimensional semiconductor structures are in focus of research for several decades.

Recently, a special interest has formed to collective and coherent phenomena in the systems with two-dimensional excitons~\cite{butov_bec,Snoke:2002aa,Gorbunov:2006aa}. In asymmetric single or double quantum well structures photogenerated electrons and holes are separated in the real space, which strongly reduces an overlap of the electron and hole wavefunctions without significant reduction of the exciton binding energy. Such excitons are termed as \emph{indirect} or \emph{dipolar}. The spatial separation of electrons and holes results in long lifetimes of the excitons and allows one to observe a number of fascinating effects, e.g., extended spatial coherence~\cite{High:2012fk}.
Measurements of transport properties of neutral excitons require, as a rule, elaborate experiments~\cite{2dexciton_tr:88,2dexciton_tr:95,PhysRevLett.99.047602}. The transport effects can be  accessed in the structures with indirect excitons, recent works have revealed nontrivial spin patterns and spin transport~\cite{doi:10.1021/nl9024227,PhysRevLett.110.246403} as well as intricate interplay of excitonic drift and diffusion~\cite{PhysRevB.91.205424}. 

The key parameter governing transport effects is the diffusion coefficient. It is determined by the properties of random potential experienced by excitons due to quantum well structure imperfections, impurities, etc., excitonic density of states and their distribution function. For excitons at low enough temperatures where exciton-phonon interaction is negligible and whose mean kinetic energy $\bar \varepsilon$ exceeds by far the collisional broadening $\hbar/\tau$ related with the disorder ($\tau$ is the scattering time) the dominant contribution to the diffusion constant is provided by random potential scattering. This classical contribution to the diffusion coefficient can be estimated as $D_{cl} \sim \bar \varepsilon \tau/M$, where $M$ is the exciton effective mass in the quantum well plane and difference between quantum (out-scattering) and momentum (or transport) relaxation times is disregarded. The interference of different classical trajectories of exciton provides the quantum correction to the classical value of diffusion coefficient, $\delta D \sim (\hbar/M) \ln(\tau/\tau^*)$ in two-dimensions. In the simplest possible case of spin-less quantum particle $\tau^*\gg \tau$ can be associated with the phase coherence time or the lifetime, whichever is shorter, hence, $\delta D$ is negative, the effect known as weak localization~\cite{Gorkov:WL,aa}. In fact, the weak localization is inherently related with the coherent backscattering of quantum particles, as demonstated for excitons in bulk semiconductors back at 1970s~\cite{1977ZhETF..72.2230I}.

It is noteworthy that quantum correction to diffusion coefficient is extremely sensitive to the fine details of the energy spectrum and, in particular, to the spin dynamics~\cite{aa,PhysRevB.77.165341}. For example, in the case of electrons, spin-$1/2$ fermions, a sufficiently strong spin-orbit coupling results in the change of sign of the quantum correction to conductivity $\delta \sigma$, which is proportional to $\delta D$, resulting in the  weak antilocalization~\cite{hikami80,iordanskii94}. In electronic systems, $\delta \sigma$ is particularly sensitive to the temperature and the external magnetic field strength~\cite{aa} making it possible to extract the values of the spin splittings, phase relaxation times, etc., see Ref.~\cite{0268-1242-24-6-064007} for review.

Excitons are composite bosons made of two fermions. Hence, on one hand, excitons demonstrate bosonic properties~\cite{butov_bec,Gorbunov:2006aa}, while on the other hand exciton dynamics is governed by its fermionic constituents. The studies of quantum correction to the exciton diffusion constant are important to elucidate the role of fermionic spin degrees of freedom in the exciton transport. The detailed theory of quantum corrections to the exciton diffusion coefficient is absent. So far, the interference of excitons  and exciton-polaritons in bulk semiconductors has been studied theoretically~\cite{1977ZhETF..72.2230I} (see also Ref.~\cite{hanamura89}) and experimentally~\cite{1977JETP...46..590G} neglecting the effect of exciton spin degrees of freedom on the interference. The theory of Refs.~\cite{PhysRevB.52.R2261,arseev98} for quantum contributions to two-dimensional exciton diffusion disregarded spin degrees of freedom as well. The spin effects in interference of exciton-polaritons in planar microcavities were studied in Refs.~\cite{PhysRevB.77.165341,PhysRevB.82.085315,PhysRevB.79.125314} taking into account just two possible spin states of optically active excitons. For indirect excitons, however, bright and dark states are close in energy and all four excitonic spin states should be taken into account. Moreover, in the systems of indirect excitons, depending on the structure parameters, the exchange interaction between the electron and the hole forming the exciton and the single-particle spin splittings caused by bulk and structure inversion asymmetries can be comparable. The theory of quantum contributions to exciton diffusion constants is proposed here. An interplay between the two-particle exchange interaction and single-particle spin splittings is analyzed in detail.  It is demonstrated that depending on the system parameters the excitons are expected to demonstrate either weak localization or weak antilocalization.

\section{Model}

\begin{figure}[t]
\includegraphics[width=\linewidth]{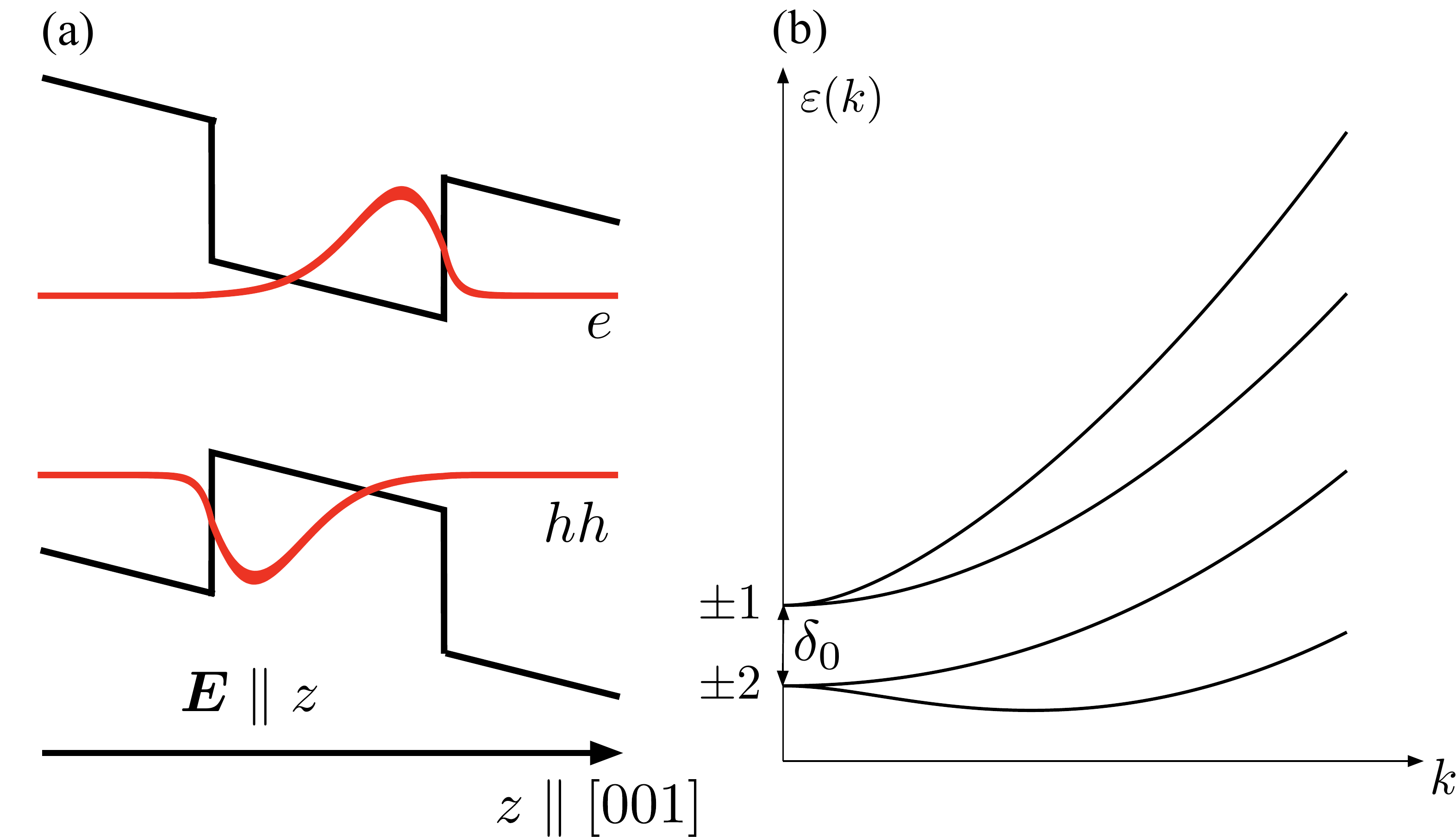}
\caption{(a) Schematic illustration of the quantum well structure in the presence of an electric field applied along the $z$-axis. (b) Sketch of the exciton dispersion calculated by diagonalization of the Hamiltonian~\eqref{Ham}.}\label{fig:scheme}
\end{figure}

We consider a two-dimensional quantum well structure grown from a zincblende lattice material where direct or indirect excitons can be formed. Figure~\ref{fig:scheme}(a) sketches a wide quantum well in the electric field applied along the growth axis $z\parallel [001]$. Depending on the electric field strength and confining potential, an overlap of an electron and hole envelope functions can be accurately controlled. The exciton is assumed to be formed of the electron occupying the ground conduction subband and the heavy hole occupying the topmost valence subband. We assume the Coulomb interaction between the electron and the hole to be strong enough to consider only the ground, $1s$-like, state of the electron-hole relative motion. As a result, the relevant exciton states are labelled by the following quantum numbers $\bm p$, the in-plane momentum of the center of mass motion, $s_e=\pm 1/2$, the electron spin-$z$ component, and $s_h=\pm 1/2$, the heavy-hole pseudospin component. The states with $s_h=\pm 1/2$ correspond to the $|\Gamma_8,\mp 3/2\rangle$ valence band Bloch functions. For such a convention, the electron and hole states with $s_e=s_h$ transform in the same way at the operations of the point groups $D_{2d}$ or $C_{2v}$ relevant for the quantum well system under consideration. The functions with $s_e=\pm 1/2$ and $s_h=\pm 1/2$ form equivalent bases of spinor representations $\Gamma_6$  or $\Gamma_5$ for point symmetry groups  $D_{2d}$ and $C_{2v}$, respectively~\cite{koster63,ivchenkopikus,PhysRevB.89.075430}. 

Following the method of invariants we represent excitonic Hamiltonian in the form~\cite{ivchenko05a,averkiev:15582,PhysRevB.89.075430}
\begin{multline}
\label{Ham}
\mathcal H(\bm p) = \frac{p^2}{2M} + \sum_{c=e,h} \bigl[(\alpha_c + \beta_c) \sigma_x^{(c)} p_y - (\alpha_c - \beta_c) \sigma_y^{(c)} p_x\bigr] + \\
+ \mathcal H_{exch}\:.
\end{multline}
Here we took into account only lowest possible powers of the momentum components, introduced the in-plane axes $x\parallel [1\bar 10]$, $y\parallel [110]$, the spin-$1/2$ and pseudospin-$1/2$ basic Pauli matrices $\sigma^{(e)}_i$ and $\sigma^{(h)}_i$ ($i=x,y,z$) acting in the space of corresponding basic spinors. Unit matrices are omitted. The exciton translational motion mass is $M$, the parameters $\alpha_c$ ($\beta_c$) describe the contributions of the structure (bulk) inversion asymmetry to the individual carrier spin splitting, $c=e$ for the electron and $c=h$ for the hole. In Eq.~\eqref{Ham} $\mathcal H_{exch}$ stands for the Hamiltonian of the exchange interaction between an electron and a hole in the exciton, 
\begin{equation}
\label{H:exc}
\mathcal H_{exch} = \frac{\delta_0}{2} \sigma^{(e)}_z  \sigma^{(h)}_z\:,
\end{equation} 
where we took into account only momentum-independent contributions. The exchange interaction~\eqref{H:exc} leads to the splitting between optically active (bright) and dipole-forbidden (dark) doublets. In GaAs-based structures $\delta_0>0$, i.e., the bright states have higher energy. In Eq.~\eqref{H:exc} we disregard the contributions $\delta_x \sigma_x^{(e)} \sigma_x^{(h)} + \delta_y \sigma_y^{(e)} \sigma_y^{(h)}$ describing the fine structure of the doublets, since, as a rule, $|\delta_{x,y}| \ll \delta_0$. It is worth mentioning that the same form of the spin splittings for electrons and holes in Eq.~\eqref{Ham} as well as the sign of $\delta_0$ term in Eq.~\eqref{H:exc} are related with the choice of the pseudospin represention for the hole states. We stress that linear-in-the-momentum terms in the heavy-hole Hamiltonian can be substantial and even exceed by far those for the electron~\cite{Rashba1988175,PhysRevLett.104.066405,PhysRevB.89.075430}; additional contribution to $\bm p$-linear terms for exciton may result from cubic-in-the-momentum terms in the heavy-hole Hamiltonian averaged over the relative motion wavefunction~\cite{PhysRevLett.110.246404}. 

The exciton Hamiltonian~\eqref{Ham} is a matrix in the space of exciton Bloch functions $|s_e,s_h\rangle$. It can be also conveniently represented as a $4\times 4$ matrix in the basis of states $|m\rangle$ with $m=s_e -3s_h$ being the angular momentum $z$-component for exciton Bloch function. The pair with $m=\pm1$ corresponds to the bright states and $m=\pm 2$ corresponds to the dark doublet. Exciton energy spectrum fine structure is determined by the competition of the $\bm p$-linear spin-orbit terms with the  exchange interaction $\mathcal H_{exch}$. Figure~\ref{fig:scheme}(b) shows the exciton dispersion and splittings of exciton spin states in the case of dominant structure inversion asymmetry with $\beta_e=\beta_h=0$, $\alpha_e,\alpha_h\ne0$. For small momenta, $|\alpha_c p| \ll \delta_0$ ($c=e,h$), the splittings of dark and bright doublets are quadratic in $\bm p$, an increase of momenta or decrease of $\delta_0$ makes spin splittings $\bm p$-linear.

The theory of excitonic coherent transport is developed by means of Greens function technique. We assume that the exciton propagates in a weak random short-range potential $U(\bm r)$ characterized by the correlation function $\langle  U(\bm r_1) U(\bm r_2)\rangle = \hbar^3/(M\tau)\delta(\bm r_1 - \bm r_2)$ where $\bm r$ is the in-plane center of mass coordinate of exciton, $\tau$ is the elastic scattering time. In our model the excitons are non-degenerate with the distribution function $f(\varepsilon) = \exp{[(\mu - \varepsilon)/k_B T]}$ characterized by the chemical potential $\mu<0$ and the  temperature $T$. Further calculations are carried out under the standard assumptions: $k_B T \tau/\hbar \gg 1$, and $\tau \ll \tau^*$, where $\tau^*$ is the effective coherence time given by $1/\tau^* = 1/\tau_0 + 1/\tau_\phi$ with exciton lifetime $\tau_0$ and dephasing time $\tau_\phi$. Additional assumptions $\delta_0, |\alpha_c p_T|, |\beta_c  p_T| \ll \hbar/\tau$, where $p_T = \sqrt{2Mk_B T}$ are necessary to derive analytical expressions, however in the numeric simulation these assumptions are relaxed~\cite{Note1}. The experimental values $D_{cl} \approx 5$~cm$^2$/s, $T = 8$~K measured in the system of indirect excitons in Ref.~\onlinecite{doi:10.1021/nl9024227} and $\beta_e \approx 5$~meV\AA~\cite{PhysRevB.86.195309} yield $\tau \approx 2.5$~ps, $k_B T \tau/\hbar \approx 2.5$, and $|\beta_e  p_T| \approx 0.2$. The retarded and advanced Greens functions averaged over the disorder in the momentum representation are given by standard expressions~\cite{iordanskii94,PhysRevB.77.165341}
\begin{equation}
\label{greens}
\mathcal G^{R,A}(\varepsilon,\bm p) = [\varepsilon - \mathcal H(\bm p) \pm \mathrm i \hbar/2\tau \pm \mathrm i \hbar/2\tau^*]^{-1}.
\end{equation}
Similarly to the Hamiltonian~\eqref{Ham} Greens functions are the matrices in the space of four excitonic states $|m\rangle$. We introduce the $16\times 16$ propagator matrix $\mathcal P(\bm q)$ with the elements (c.f. Ref.~\cite{golub05}):
\begin{multline}
\label{prop}
[\mathcal P(\bm q)]_{mm_1}^{m'm_1'} = \frac{\hbar^3}{M\tau} \sum_{\bm p} \mathcal G^{R}_{mm'}(\varepsilon,\bm p) \mathcal G^{A}_{m_1m_1'}(\varepsilon,\bm q- \bm p) = \\
%\frac{\hbar}{\tau} \int_0^{2\pi} \frac{d\varphi_{\bm p}}{2\pi} \int \frac{dE_p}{2\pi} \mathcal G^{R}_{mm'}(\varepsilon,\bm p) \mathcal G^{A}_{m_1m_1'}(\varepsilon,\bm q- \bm p),
\frac{1}{1+\tau/\tau^*}
\int \limits_0^{2\pi} \frac{d\varphi_{\bm p}}{2\pi} \left[1 + \mathrm i q l_0 \cos{(\varphi_{\bm q} - \varphi_{\bm p})} + \mathrm i \tau \mathcal V(\bm p) \right]^{-1}\:,
\end{multline}
where the matrix $\mathcal V(\bm p)$ is given by
\begin{equation}
\mathcal V (\bm p) = \bm L_e \bm \Omega_e + \bm L_h \bm \Omega_h + 
\frac{\delta_0}{2} \left[ \sigma_z^{(e),1}\sigma_z^{(h),1} - \sigma_z^{(e),2}\sigma_z^{(h),2} \right]\:.
\end{equation}
Here $l_0 = p_0\tau/M$ is the mean free path, $p_0 = \sqrt{2M\varepsilon_0}$, $\bm L_c = [\bm \sigma^{(c),1} + \bm \sigma^{(c),2}]/2$, superscripts $1$ and $2$ refer to the spin operators on the time-reversed trajectories~\cite{iordanskii94}, and the components of pseudovectors $\bm \Omega_e$, $\bm \Omega_h$ are given by $\Omega_{c,x} = 2 \alpha_c p_0\sin{\varphi_{\bm p}}/\hbar$, $\Omega_{c,y} = -2 \alpha_c p_0\cos{\varphi_{\bm p}}/\hbar$. The integral over $|\bm p|$ in Eq.~\ref{prop} is taken using residue theorem making use of the fact that for typical exciton energies $E_p \sim k_B T$, where $E_p=p^2/2M$ is the exciton dispersion, the product $E_p \tau/\hbar$ exceeds unity by far.
We further introduce the Cooperon matrix
\begin{equation}
\label{Cooperon}
\mathcal C(\bm q) = [{1-\mathcal P(\bm q)}]^{-1},
\end{equation}
which describes the interference of excitonic waves propagating in disordered quantum well as a result of coherent backscattering, it represents the sum of ``maximally crossed'' diagrams~\cite{1977ZhETF..72.2230I,Gorkov:WL,aa}.
%In Eq.~\eqref{prop} $E_p=p^2/2M$ is the exciton dispersion and integral over the exciton energy should be taken using residue theorem making use of the fact that for typical exciton energies $E_p \sim k_B T$ the product $E_p \tau/\hbar$ exceeds unity by far. The normalization area is set to unity hereafter. 

The diffusion coefficient for the non-degenerate exciton gas  can be conveniently expressed as
\begin{equation}
\label{D:gen}
D = \frac{1}{M k_B T}\int_0^\infty d\varepsilon \exp \left( -\frac{\varepsilon}{k_BT} \right) \left[ \varepsilon \tau + \hbar W(\varepsilon)\right],
\end{equation}
where the first term in the square brackets describes the classical contribution to the diffusion coefficient $D_{cl} = k_B T\tau/M$ and the second term is the leading order quantum correction
\begin{equation}
\label{W:gen}
W(\varepsilon)= - \sum_{\bm q,m,m'} \mathcal C_{mm'}^{m'm}(\bm q).
\end{equation}
Equations~\eqref{Cooperon} and \eqref{W:gen} are derived in the diffusion approximation where the logarithms $\ln(\tau^*/\tau)$ and $\ln(\tau_s/\tau)$ ($\tau_s$ is the typical spin relaxation time, see below for details) are assumed to be large.

\section{Results and discussion}

Equation~\eqref{D:gen} shows that the interference contribution to the diffusion coefficient can be calculated for the monoenergetic excitons and then averaged over the distribution function. Qualitatively, this is because the interference processes can be interpreted as a modification of the exciton scattering cross-section at a given defect~\cite{dmitriev97,PhysRevB.77.165341,PhysRevB.82.085315}. Therefore, we focus on the discussion of the quantity $W$ in Eqs.~\eqref{D:gen}, \eqref{W:gen}, which controls the interference correction for a fixed value of $\varepsilon \equiv \varepsilon_0$ ($\varepsilon_0 \tau/\hbar \gg 1$). Moreover, for simplicity we assume that the spin splittings for both electrons and holes are dominated by one mechanism only, e.g., $\alpha_c\ne 0$, $\beta_c=0$~\cite{Note2}. Below we address several limits which demonstrate different regimes of interference of excitons. 

It is instructive to start with the limit of absent exchange interaction, $\delta_0=0$. 
%\MD{\emph{(All the parameters are already discussed under Eq.(4))}}
%The matrix $\mathcal P(\bm q)$ can be conveniently represented as 
%\begin{multline}
%\label{P:no:exch}
%\mathcal P(\bm q) = \frac{1}{1+\tau/\tau^*} \times\\
%\int \limits_0^{2\pi} \frac{d\varphi_{\bm p}}{2\pi} \left[1 + \mathrm i q l_0 \cos{(\varphi_{\bm q} - \varphi_{\bm p})} + \mathrm i \tau( \bm L_e \bm \Omega_e + \bm L_h \bm \Omega_h) \right]^{-1},
%\end{multline}
%where $l_0 = p_0\tau/M$ is the mean free path, $p_0 = \sqrt{2M\varepsilon_0}$, $\bm L_c = [\bm \sigma^{(c),1} + \bm \sigma^{(c),2}]/2$, superscripts $1$ and $2$ refer to the spin operators on the time-reversed trajectories~\cite{iordanskii94}, and the components of pseudovectors $\bm \Omega_e$, $\bm \Omega_h$ are given by $\Omega_{c,x} = 2 \alpha_c p_0\sin{\varphi_{\bm p}}/\hbar$, $\Omega_{c,y} = -2 \alpha_c p_0\cos{\varphi_{\bm p}}/\hbar$. 
Decomposing {Eq.~\eqref{prop}} in the series in small parameters $ql_0, \Omega_c \tau, \tau/\tau^* \ll 1$, substituting $\mathcal P(\bm q)$ into Eq.~\eqref{Cooperon}, then $\mathcal C(\bm q)$ into Eq.~\eqref{W:gen} and making standard transformations we arrive at the following logarithmic contributions to the interference amplitude
\begin{widetext}
\begin{multline}
\label{W:no:exch}
W= \ln\left(\frac{ \tau}{\tau^*}\right) + 3 \ln \left( \frac{\tau}{\tau^*} + \frac{\tau}{2\tau_{e}} + \frac{\tau}{2\tau_h} \right) - 2\ln \left( \frac{\tau}{\tau^*} + \frac{\tau}{2\tau_{e}} \right) - 2\ln \left( \frac{\tau}{\tau^*} + \frac{\tau}{2\tau_{h}} \right) \\
- \ln \left( \frac{\tau}{\tau^*} + \frac{\tau}{\tau_{e}} \right) - \ln \left( \frac{\tau}{\tau^*} + \frac{\tau}{\tau_{h}} \right) + 2\ln \left( \frac{\tau}{\tau^*} + \frac{3\tau}{4\tau_{e}} + \frac{3\tau}{4\tau_{h}} + \frac{\tau}{\tau_1}  \right) +2\ln \left( \frac{\tau}{\tau^*} + \frac{3\tau}{4\tau_{e}} + \frac{3\tau}{4\tau_{h}} - \frac{\tau}{\tau_1}  \right)
\end{multline}
\[
+ \ln \left( \frac{\tau}{\tau^*} + \frac{3\tau}{4\tau_{e}} + \frac{3\tau}{4\tau_{h}} + \frac{\tau}{\tau_2}  \right) +\ln \left( \frac{\tau}{\tau^*} + \frac{3\tau}{4\tau_{e}} + \frac{3\tau}{4\tau_{h}} - \frac{\tau}{\tau_2}  \right) \:,
\]
where $\tau_c = 1/(\Omega_c^2\tau)$ is the spin $z$ component of the electron-in-exciton ($c=e$) or the hole-in-exciton ($c=h$) relaxation time, 
\end{widetext}
\[
\frac{1}{\tau_1} = \frac14 \sqrt{\frac{1}{\tau_e^{2}} + \frac{1}{\tau_h^{2}} + \frac{14}{\tau_e\tau_h}} \:, \quad
\frac{1}{\tau_2} = \frac14 \sqrt{\frac{1}{\tau_e^{2}} + \frac{1}{\tau_h^{2}} + \frac{34}{\tau_e\tau_h}} .
\]

Equation~\eqref{W:no:exch} shows that the sign of interference contribution to the excitonic diffusion coefficient can be positive or negative depending on particular relation between the characteristic times $\tau^*$, $\tau_e$, $\tau_h$. Although in the considered limit the spin states of the electron and the hole are not mixed, 
$\delta_0=0$, the correlation of the electron and hole motion in the exciton does not allow one to reduce the interference effect to the combination of the electron and hole contributions. Only if $\alpha_e\alpha_h=0$, i.e. the spin splitting for one of the carriers is absent, the interference correction to the diffusion coefficient reduces to the twice single-carrier contribution:
\begin{equation}
\label{omegah_zero}
W =  -2\ln \left(\frac{\tau}{\tau^*}\right) + 4\ln \left( \frac{\tau}{\tau^*} + \frac{\tau}{2\tau_{s}} \right) + 2\ln \left( \frac{\tau}{\tau^*} + \frac{\tau}{\tau_{s}} \right) \:.
\end{equation}
Here $\tau_s$ is the spin relaxation time of the carrier with non-zero spin splitting. In such a case the antilocalization behavior is expected for sufficiently strong spin-orbit splitting: For $\tau_s \ll \tau^*$ the interference amplitude $W>0$ and it demonstrates non-monotonic behavior with a decrease of $\tau^*$ (i.e. caused by the temperature increase). This is because a decrease in $\tau^*$ results, firstly, in the suppression of positive contribution, $\propto \ln(\tau/\tau^*)$, and only at $\tau^* \sim \tau_s$ the negative contributions in Eq.~\eqref{omegah_zero} become suppressed~\cite{hikami80,iordanskii94}. By contrast, if the spin splittings for the electron and the hole are the same, $\alpha_e=\alpha_h$, the interference contribution remains negative and increases monotonously with a decrease in  $\tau^*$:
\begin{multline}
\label{omega_eq}
W = 2\ln{\left(\frac{\tau}{\tau^*}\right)} + \ln \left( \frac{\tau}{\tau^*} + \frac{\tau}{\tau_e} \right) - 2 \ln \left( \frac{\tau}{\tau^*} + \frac{\tau}{2\tau_e} \right)\\
+ 2\ln \left( \frac{\tau}{\tau^*} + \frac{5\tau}{2\tau_{e}} \right) + \ln \left( \frac{\tau}{\tau^*} + \frac{3 \tau}{\tau_{e}} \right).
\end{multline}
Indeed, for $\alpha_e = \alpha_h$ the exciton spin Hamiltonian can be conveniently expressed as $\hbar (\bm S\cdot \bm \Omega_e)$ describing interaction of the total spin of the electron and the hole, $\bm S = (\bm \sigma^{(e)} + \bm \sigma^{(h)})/2$, with an effective momentum-dependent magnetic field. Since formally $S=0$ or $1$, the  phase of an exciton wavefunction acquired while exciton propagates along the close loop amounts to the integer number of $2\pi$ and the interference is always constructive. 

\begin{figure*}[t]
\includegraphics[width=0.97\textwidth]{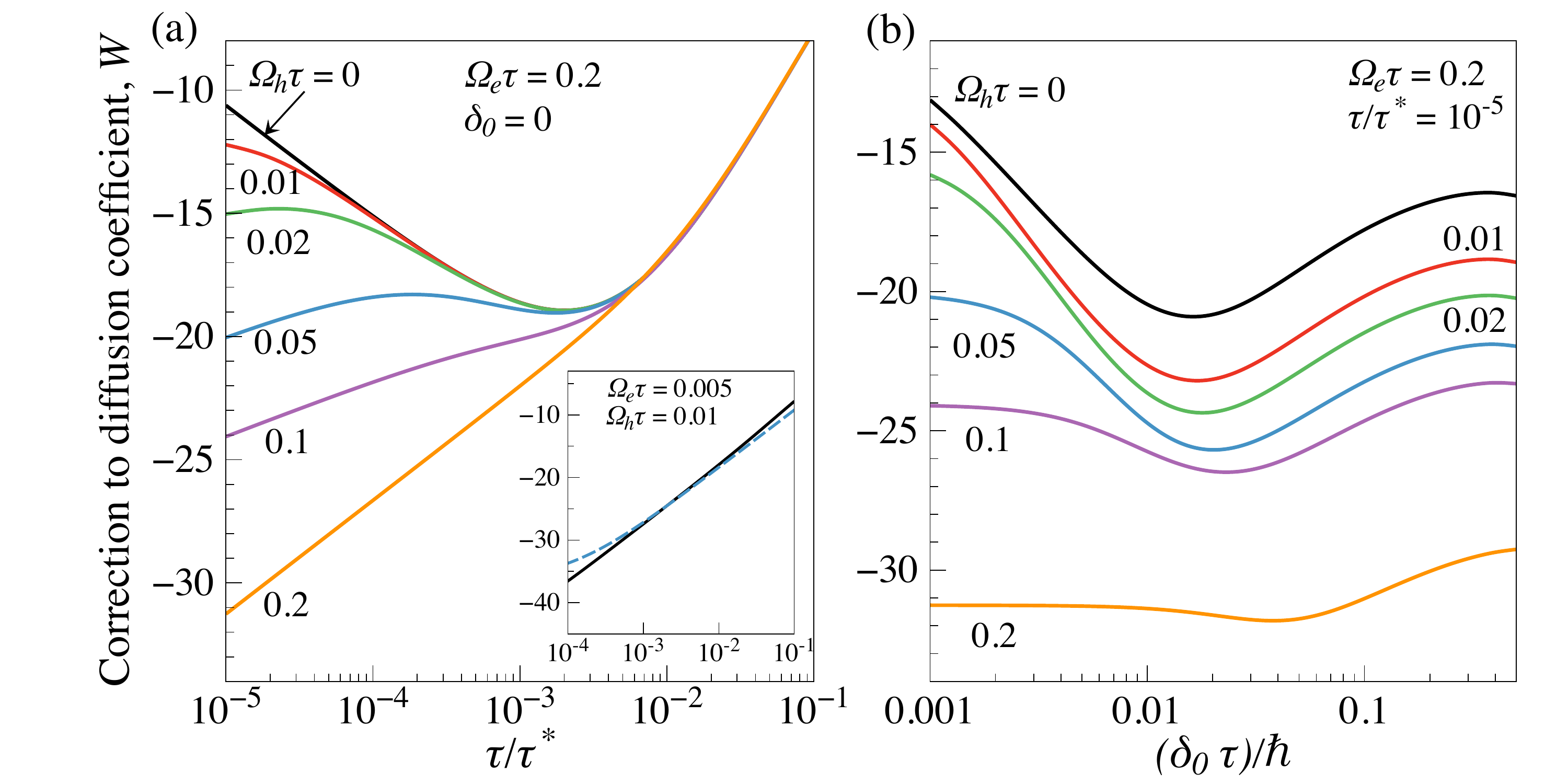}
\caption{Interference contribution to the diffusion coefficient of excitons. (a) Results of the calculation neglecting exchange interaction, $\delta_0=0$, presented for different values of $\Omega_h \tau$ ($\Omega_h \tau = 0,~0.01,~0.02,~0.05,~0.1,~0.2$) at a fixed $\Omega_e \tau = 0.2$ ($\Omega_{c} = 2 |\alpha_{c} p_0|/\hbar$). The inset shows comparison of results of numerical calculations (solid curve) and analytical calculations (dashed curve) using Eq.~\eqref{W:no:exch} for $\Omega_e\tau = 0.005$, $\Omega_h \tau = 0.01$. (b) Interference contribution as a function of the exchange splitting $\delta_0\tau/\hbar$ presented for different values of $\Omega_h \tau$ ($\Omega_h \tau = 0,~0.01,~0.02,~0.05,~0.1,~0.2$) at fixed $\Omega_e \tau = 0.2$ and $\tau/\tau^* = 10^{-5}$.} \label{fig:fig2}
\end{figure*}

Now we switch to the opposite limit of a very strong exchange interaction $\delta_0 \gg \hbar\Omega_e,\hbar\Omega_h$. In this case bright and dark doublets are well separated in energy and the spin-orbit splittings of the conduction and valence bands can be accounted for by means of second order perturbation theory. The effective spin-Hamiltonians $\mathcal H_{1,2}$ of the doublets have, in agreement with the symmetry arguments~\cite{ivchenko05a}, the form
\begin{equation}
\label{H:large:split}
\mathcal H_1 = \begin{pmatrix}
\delta_0/2 & \hbar \Omega_1 \mathrm e^{2\mathrm i \varphi_{\bm p}}\\
\hbar \Omega_1 \mathrm e^{-2\mathrm i \varphi_{\bm p}}  &\delta_0/2 
\end{pmatrix},~ 
\mathcal H_2 = -\begin{pmatrix}
\delta_0/2 & \hbar \Omega_1 \\
\hbar \Omega_1  &\delta_0/2 
\end{pmatrix},
\end{equation}
with $\Omega_1 = -\hbar \Omega_e\Omega_h/(2\delta_0)$ describing the splittings of bright and dark doublets. Note that the form of the bright excitons Hamiltonian $\mathcal H_1$ is similar to that of the exciton-polaritons in planar microcavities~\cite{kavokin05prl,Leyder:2007ve,PhysRevB.77.165341}. Hence, in the limit of strong exchange splitting an interference of doublets is independent and the transport of excitons is provided by two independent channels. For the interference amplitude $W$ we obtain
\begin{equation}
\label{W:large:split}
W = 3 \ln\left(\frac{\tau}{\tau^*}\right) + 2 \ln \left( \frac{\tau}{\tau^*} + \frac{\tau}{2\tau_{s}} \right) - \ln \left( \frac{\tau}{\tau^*} + \frac{\tau}{\tau_{s}} \right)\:,
\end{equation}
where $\tau_s = 1/(\Omega_1^2\tau) \gg \tau$ is the relaxation rate of $z$-pseudospin component of the bright doublet~\cite{PhysRevB.77.165341}. The spin splitting of dark states is wavevector independent and is not manifested in the interference contribution, Eq.~\eqref{W:large:split}. One can see that exciton interference results in the weak localization, $W<0$. This is because dark states contribute as spin-less particles and the bright states experience effective field which is described by the second angular harmonics of $\varphi_{\bm p}$ (compared with first angular harmonics in the case of electrons or holes). Hence, even if the spin-orbit coupling is sufficiently strong, the phase acquired by the exciton wavefunction while travelling around the closed loop is given by $2\pi$ rather then $\pi$ for electrons~\cite{PhysRevB.77.165341}.

An interplay of single particle spin splittings and exchange interaction in the  interference contribution to the exciton diffusion coefficient is illustrated in Fig.~\ref{fig:fig2}. By contrast to the analytical results presented above in Eqs.~\eqref{W:no:exch}, \eqref{omegah_zero}, \eqref{omega_eq}, and \eqref{W:large:split} with a logarithmic accuracy, i.e., retaining only the leading terms in $\ln(\tau^*/\tau)$, $\ln(\tau_s/\tau)$, in numerical calculations shown in Fig.~\ref{fig:fig2} the non-logarithmic [cf. Ref.~\cite{PhysRevB.70.155323}] terms are also included. To that end, in numerical procedure, we include in Cooperon only contributions due to three and more scattering processes presenting $\mathcal C(\bm q) = \mathcal P^3 (\bm q)[{1-\mathcal P(\bm q)}]^{-1}$ and numerically integrating in Eq.~\eqref{W:gen} over $|\bm q|$ varying from zero to infinity~\cite{Note3}. In the limit $\Omega_c \tau \ll 1$, when logarithmic terms are large, the numerical procedure yields results, which are close to the ones given by Eq.~\eqref{W:no:exch}, compare solid and dashed lines in the inset to Fig.~\ref{fig:fig2}(a). However for the parameters used in the calculation of the curves in the main panels the non-logarithmic contributions~\cite{PhysRevB.70.155323} are not negligible and affect both the amplitude and the shape of the $W$ curves. The limit of absent exchange interaction, $\delta_0=0$, is depicted in Fig.~\ref{fig:fig2}(a) for different values of $\alpha_h$ and a fixed value of $\alpha_e$. As it was discussed above [see Eqs.~\eqref{omegah_zero} and \eqref{omega_eq}], increasing $\alpha_h$ from $\alpha_h = 0$ to $\alpha_h = \alpha_e$ results at $\tau^* \gg \tau_e$ in the transition from weak antilocalization regime (corresponding to a negative slope of the $W$ dependence on $\tau/\tau^*$) to weak localization regime (positive slope of a $W$ curve). Note that, in contrast to the similar dependences for electrons~\cite{golub05}, the curves presented in Fig.~\ref{fig:fig2}(a) have two extrema for the intermediate values of $\Omega_h$ ($\Omega_h \tau = 0.01$, 0.02, 0.05). Figure~\ref{fig:fig2}(b) presents the dependence of $W$ on the exchange strength $\delta_0 \tau/\hbar$ at a fixed value $\tau^*/\tau = 10^5$. Increasing the exchange splitting between the bright and dark exciton doublets, in accordance with Eq.~\eqref{W:large:split}, results in negative correction to diffusion coefficient for all values of $\alpha_h$. In the limit $\delta_0 \tau/\hbar \gg 1$ (not shown) all the curves saturate at the value $W = 4 \ln(\tau/\tau^*)$ [see Eq.~\eqref{W:large:split}], which corresponds to the coherent backscattering in four independent transport channels.

\section{Conclusion}

The theory of interference contribution to the exciton diffusion coefficient in quantum wells has been put forward. An interplay of four spin states arising from $\pm 1/2$ electron spin and $\pm 3/2$ heavy hole spin has been taken into account. An interplay of the single particle spin-orbit splittings and the electron-hole exchange interaction has been shown to result in weak localization or antilocalization of excitons. Particularly, if exchange interaction dominates over the spin splittings or if spin splittings of the electron and the hole are the same, excitons demonstrate weak localization. By contrast, if the main contribution to the exciton energy spectrum is provided by one of the charge carriers, the coherent backscattering results in the destructive interference of excitons and their weak antilocalization takes place. The interference effects can be accessed by the transport experiments similar to those in Refs.~\cite{2dexciton_tr:88,2dexciton_tr:95,PhysRevLett.110.246403,PhysRevB.91.205424}. The magnitude of the interference contribution is sensitive to the temperature via the coherence time $\tau^*$ and the distribution function of excitons, and to the magnetic field which affects exciton interference due to flux of the field ``passing through'' the exciton~\cite{arseev98}. The sensitivity of the interference contribution to the exciton diffusion coefficient to the spin-orbit splittings and the electron-hole exchange interaction may be also useful to study the Berry phase effects in excitonic transport~\cite{PhysRevLett.101.106401}.

\acknowledgements

Authors are grateful to L.E. Golub for fruitful discussions. This work was partially supported by RFBR, Dynasty foundation and RF President Grant MD-5726.2015.2.

\end{document}